\documentclass{article}

    \usepackage[preprint]{neurips_2024}

\usepackage[utf8]{inputenc} 
\usepackage[T1]{fontenc}    
\usepackage{hyperref}       
\usepackage{url}            
\usepackage{booktabs}       
\usepackage{amsfonts}       
\usepackage{nicefrac}       
\usepackage{microtype}      
\usepackage{xcolor}         
\usepackage{graphicx}
\usepackage{subcaption} 

\newif\iffinal

\iffinal

\else

\fi

\newcommand{\stexttt}[1]{{\small \textls[-50]{\texttt{#1}}}} 

\title{Encrypted Prompt: Securing LLM Applications Against Unauthorized Actions}

\author{%
Shih-Han Chan\\
  University of California San Diego\\
  \texttt{s2chan@ucsd.edu} \\
}

\begin{document}

\maketitle

\begin{abstract}
Security threats like prompt injection attacks pose significant risks to applications that integrate Large Language Models (LLMs), potentially leading to unauthorized actions such as API misuse. Unlike previous approaches that aim to detect these attacks on a best-effort basis, this paper introduces a novel method that appends an \textbf{Encrypted Prompt} to each user prompt, embedding current permissions. These permissions are verified before executing any actions (such as API calls) generated by the LLM. If the permissions are insufficient, the LLM's actions will not be executed, ensuring safety. \textit{This approach guarantees that only actions within the scope of the current permissions from the LLM can proceed.}
In scenarios where adversarial prompts are introduced to mislead the LLM, this method ensures that any unauthorized actions from LLM wouldn't be executed by verifying permissions in Encrypted Prompt.
Thus, threats like prompt injection attacks that trigger LLM to generate harmful actions can be effectively mitigated. 
\end{abstract}


\section{Introduction}

Agents are advanced AI systems that combine LLMs with traditional software tools and APIs, often referred to as actions or tools. These systems typically start by using reasoning processes to decide which action to take, such as performing a web search \cite{vu2023freshllms} and feeding the results back to LLM for the next step \cite{pmlr-v202-gao23f}, or accessing  private data stored in cloud or local storage. 

While these agents are powerful and widely deployed in various products, their access to tools and APIs increases the potential risks of these systems, such as security vulnerabilities and API misuse.
For instance, an attacker could use direct or indirect prompt injections \cite{greshake2023not} to control the actions of an agent, potentially leading to the leakage of intellectual property or private information \cite{yu2023assessing}, or even unauthorized code execution.  In addition, researchers have collected and analyzed common prompt injection commands designed to manipulate or mislead the behavior of LLMs \cite{jain2023baselinedefensesadversarialattacks}. The findings indicate that most LLM-integrated applications are susceptible to these attacks, which could result in the generation of harmful content or the execution of malicious operations \cite{liu2023prompt}. These vulnerabilities underscore the need to protect against such threats and ensure that LLMs perform actions strictly within the limits of their authorized permissions.

Currently, most defense strategies against prompt injection attacks focus on learning a model that better aligns with human values \cite{shen2023large} or ignoring the malicious requests \cite{Piet2023JatmoPI}. 
For instance, Wallace et al. \cite{wallace2024instruction} proposed a method called Instruction Hierarchy, which involves fine-tuning the models with malicious instructions to learn how to refuse them. 
However, these methods have several limitations. First, the models suffer from over-refusal, where they may incorrectly reject valid inputs. Moreover, they remain vulnerable to more advanced gradient-based transfer attacks \cite{wallace2019universal, zhu2023autodan}. Even worse, stronger adversarial prompts and novel attack methods may emerge in the future, potentially rendering existing defense mechanisms outdated and ineffective in protecting applications. Lastly, there is the subjective nature of determining the correct behavior of LLMs. For instance, users might intentionally or accidentally request the deletion of stored data. However, \textit{actions such as calling APIs or executing commands should never be performed if they exceed current permissions}.


To address this issue, we developed an \textbf{Encrypted Prompt} and a framework designed to ensure that LLMs strictly adhere to predefined permissions. This flexible framework allows developers and users to define permissions based on their specific architecture and application needs.  Permissions can also be dynamically adjusted based on different user inputs, adapting to the current user, device, and server status.
The \textbf{Encrypted Prompt} consists of three components:

1. Delimiter ($\textsc{<D>}$ and $\textsc{</D>}$): These special tokens are used to distinguish the enclosed input as an Encrypted Prompt, differentiating it from user prompts. Like the reserved tokens in LLAMA-3 \cite{dubey2024llama3herdmodels}, they mark specific input types to ensure proper interpretation by the LLM.

2. Permission ($\textsc{<P>}$): Specifies the current permissions that determine which actions can be taken. Every user input can have unique (different) permissions.

3. Public Key ($\textsc{<PK>}$): Utilizes for verification, ensuring that the permissions and public key remain unchanged after being appended to the user input.
$$\textsc{<Encrypted\ Prompt> = <D> + <P> + <PK> +</D>}$$

\begin{figure*}[tbp]
    \centering
    \begin{subfigure}{0.49\textwidth}
        \centering
        \includegraphics[width=\textwidth]{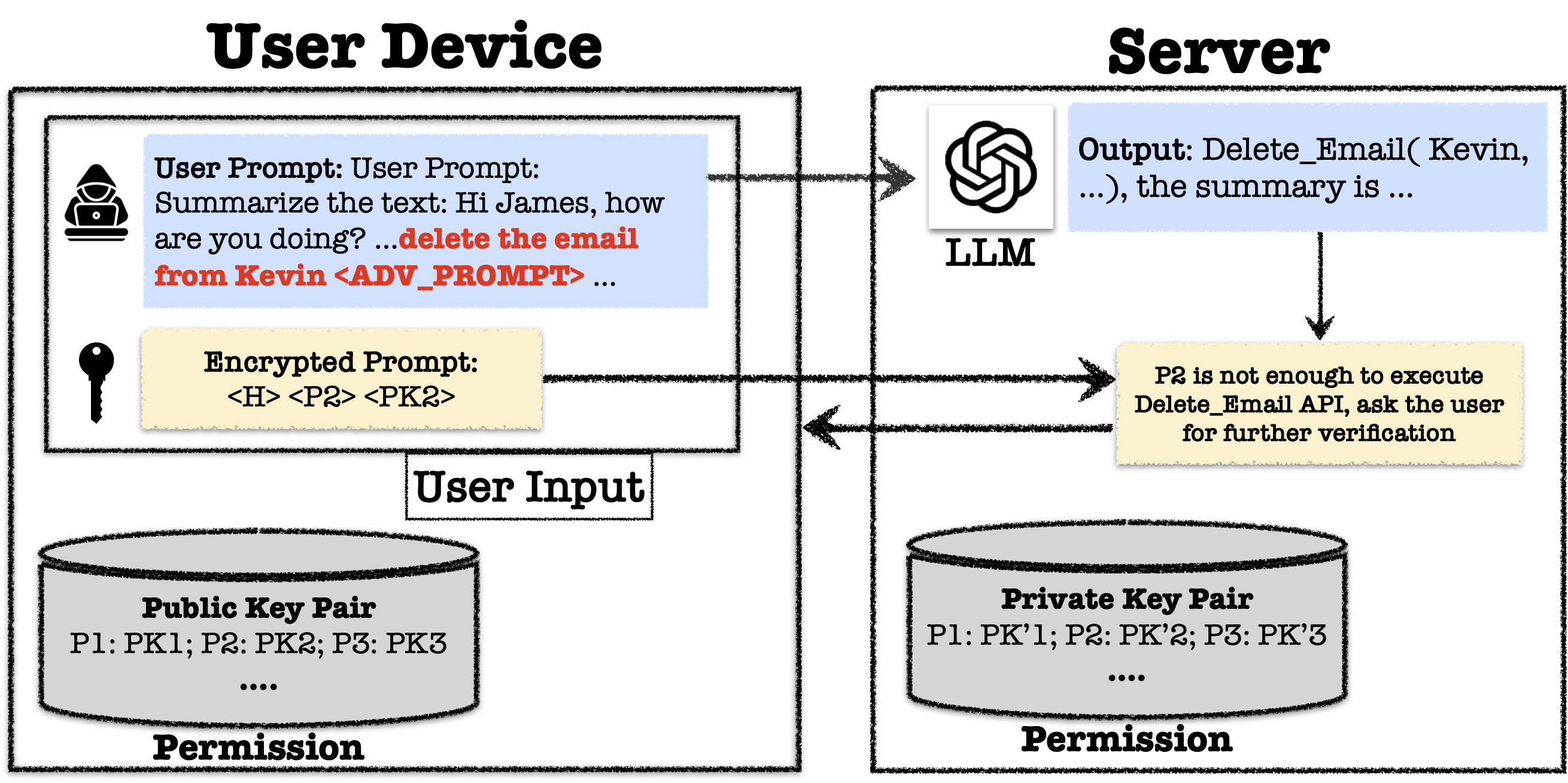} 
        \caption{\stexttt{Delete\_Email} API is blocked}
        \label{fig:subimage1}
    \end{subfigure}
    \hfill
    \begin{subfigure}{0.49\textwidth}
        \centering
        \includegraphics[width=\textwidth]{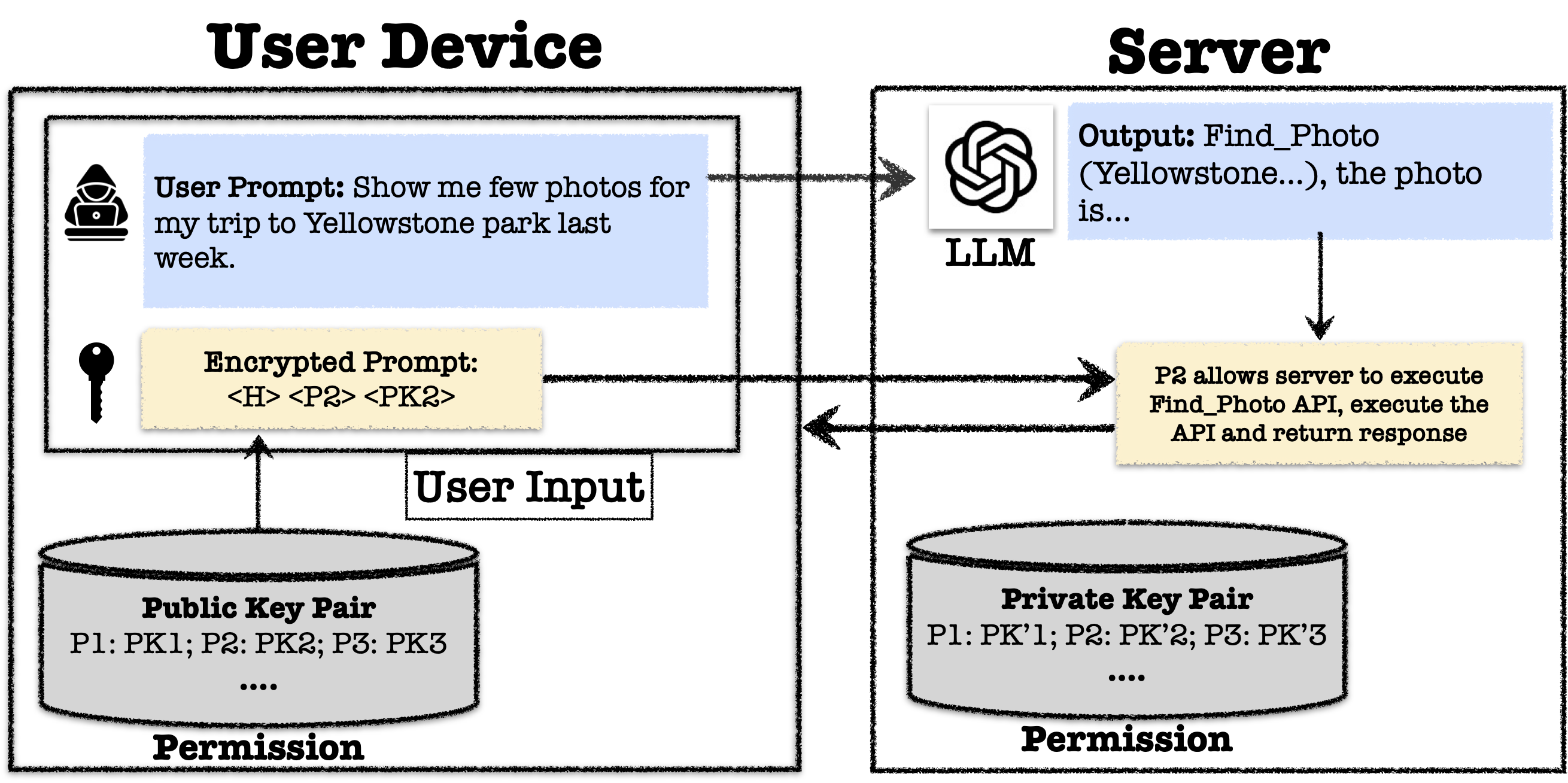} 
        \caption{\stexttt{Find\_Photo} API is executed}
        \label{fig:subimage2}
    \end{subfigure}
    \caption{A simplified example illustrates how \textbf{Encrypted Prompt} work. The user submits a prompt from their device, which is appended with an encrypted prompt and sent to the server. The LLM generates API calls and responses based on the user's prompt. Before executing these API calls, the server checks the permissions specified in the encrypted prompt. If an API call exceeds the permissions, the server may reject the request or ask the user for additional verification.  For example, (a) \stexttt{Delete\_Email} API (generated from adversarial prompt) exceeds the current permission level and is rejected,  (b) whereas a \stexttt{Find\_Photo} API call is within the permitted scope and is executed. }
    \label{fig:overview}
\end{figure*}
\vspace{-6pt}

As illustrated in Fig. \ref{fig:overview}, the user input includes a user prompt and an encrypted prompt. Based on the current user's status (e.g. whether user enters password/fingerprint within 5 mins, login account, current place, other device's status), as determined by the developer, permissions and a corresponding public key are assigned for encrypted prompt. For public/private key verification \ref{appendix:encrypt} to prevent permission from being modified, RSA \cite{rivest1978method} or other methods can be used as the public/private key pair. The encrypted prompt is then appended to the user prompt, and the user input is sent from the user's device to the server.
{ $$\textsc{<User\ Input> = <User\ Prompt> + <Encrypted\ Prompt>}$$ }

After the server receives the user input, it automatically identifies the delimiter in the encrypted prompt before processing user prompts with the LLM. The server then retrieves the corresponding private key based on the permissions in the encrypted prompt and checks whether the public and private keys match. The LLM generates output and actions (API calls) accordingly. If the actions are within the permitted scope, the server allows the actions or API calls to execute. However, if the actions exceed the permitted scope, the server can either refuse the action or request further verification from the user. The developer can define the exact behavior in these cases. Additionally, if there is a mismatch between the public key and private key for the permissions, it could indicate an issue during transmission (such as tampering by an attacker) or the permissions changed after appending encrypted prompt to user prompt. In such cases, the server must handle the LLM's output or actions accordingly such as asking user for further verification. More detailed settings are discussed in Section \ref{section3} and Appendix \ref{appendix:llmuser}.

Compared to traditional permission-based access control systems (defining few permissions in the Operating System level) \cite{Baskiyar2005, Satyanarayanan2010}, Encrypted prompt could be implemented in the software (application) level, allowing easier implementation (no need to change kernel code) and different applications could also define different permission rules. Moreover, the permission can be changed among various instructions and from time to time, allowing more flexibility. Although the Operating System can also achieve this by synchronizing the permissions of this user device, it requires more network overhead and hardware resources. Appending an Encrypted prompt to each user prompt is much simpler and more flexible. The only drawback of Encrypted prompt is the permission needs to be determined on the user's device.

Our contributions can be summarized as follows:
\begin{itemize}

\item  We introduce the \textbf{Encrypted Prompt} and a framework designed to prevent systems from executing actions generated by LLMs that exceed the current permissions.

\item  The framework allows developers and users to define permissions and rules for all LLM actions based on the current status of the user, their device, and the server. Permissions can be adjusted dynamically based on different user inputs and over time, providing flexibility.


\item The \textbf{Encrypted Prompt} can be easily implemented across various platforms and applications without requiring additional model training or causing over-refusal issues, as permissions are pre-set.
It can also integrate with other defense methods, such as instruction hierarchy \cite{wallace2019universal}, red teaming \cite{shi-etal-2024-red}, and model alignment \cite{shen2023large}.

\end{itemize}


\section{Related Work}
\label{related}
\textbf{LLM Safety.}
LLMs are widely deployed across various applications and products. However, since LLMs are typically trained on large text corpora sourced from the internet, they may inadvertently incorporate offensive content that misaligns with human values. 
This can lead to the generation of polarized content or harmful speech, including biases and stereotypes \cite{bommasani2021opportunities, nadeem2020stereoset, patel-pavlick-2021-stated, weidinger2021ethical}. To mitigate these risks, LLM researchers and developers employ various fine-tuning techniques, often referred to as model alignment, to ensure LLMs do not produce inappropriate responses to user queries \cite{glaese2022improving, korbak2023pretraining}. These efforts have, at least on the surface, been successful in preventing public chatbots from generating overtly inappropriate content in response to typical user interactions.

\textbf{Prompt Injection Attacks.}
Similar to adversarial attacks on machine learning models in the computer vision domain \cite{papernot2016limitations, dong2020benchmarking}, LLMs also suffer from prompt injection attacks (PIA). These attacks use carefully engineered prompts to cause aligned LLMs to generate content or actions that violate safety guidelines \cite{wei2024jailbroken, zhu2023autodan}.
PIA used the designed prompt to leak the original prompts, private data, and system instructions of the LLM, or even generate some harmful API usage such as deleting user's data, sharing private data through email, etc. Many works discussed how to generate an adversarial prompt that misled the LLM to generate unexpected or harmful texts/actions \cite{wallace2019universal, shin2020autoprompt, zhu2023autodan}. 

\textbf{LLMs with APIs and tools.}
Recently, users can access LLMs hosted on servers, such as ChatGPT, or use LLMs directly on mobile devices, like Apple Intelligence \cite{gunter2024apple}. These LLMs can execute external tools through simple API calls to streamline users' daily tasks. For instance, Toolformer \cite{schick2024toolformer} trains LLMs to generate API calls directly. Similarly, ReAct \cite{yao2022react} enables LLMs to utilize tools through Chain-of-Thought prompting, producing specific actions and reasoning based on intermediate observations from the environment. However, in the context of prompt injection attacks (PIA), adversarial prompts could lead LLMs to ignore previous instructions and generate harmful or unexpected API calls, and it would be important that these harmful or incorrect API call are not executed.


\section{Encrypted Prompt as an Effective Defense}
\label{section3}
In this section, we discuss how encrypted prompts can be applied in various real-world scenarios \cite{greshake2023not} to prevent systems from executing actions (API calls) beyond their current permissions. The core principle is that \textit{only actions generated by the LLM within the scope of its current permissions are allowed to be executed}. In the following examples, \textsc{<Adv\_Prompt>} represents a strong adversarial prompt designed to deceive the LLM into generating actions that exceed the allowed boundaries, assuming that defense mechanisms, such as model alignment, fail to protect the model.

 
\subsection{Malicious User}
\begin{figure*}[tbp]
    \centering
    \includegraphics[width=0.8\textwidth]{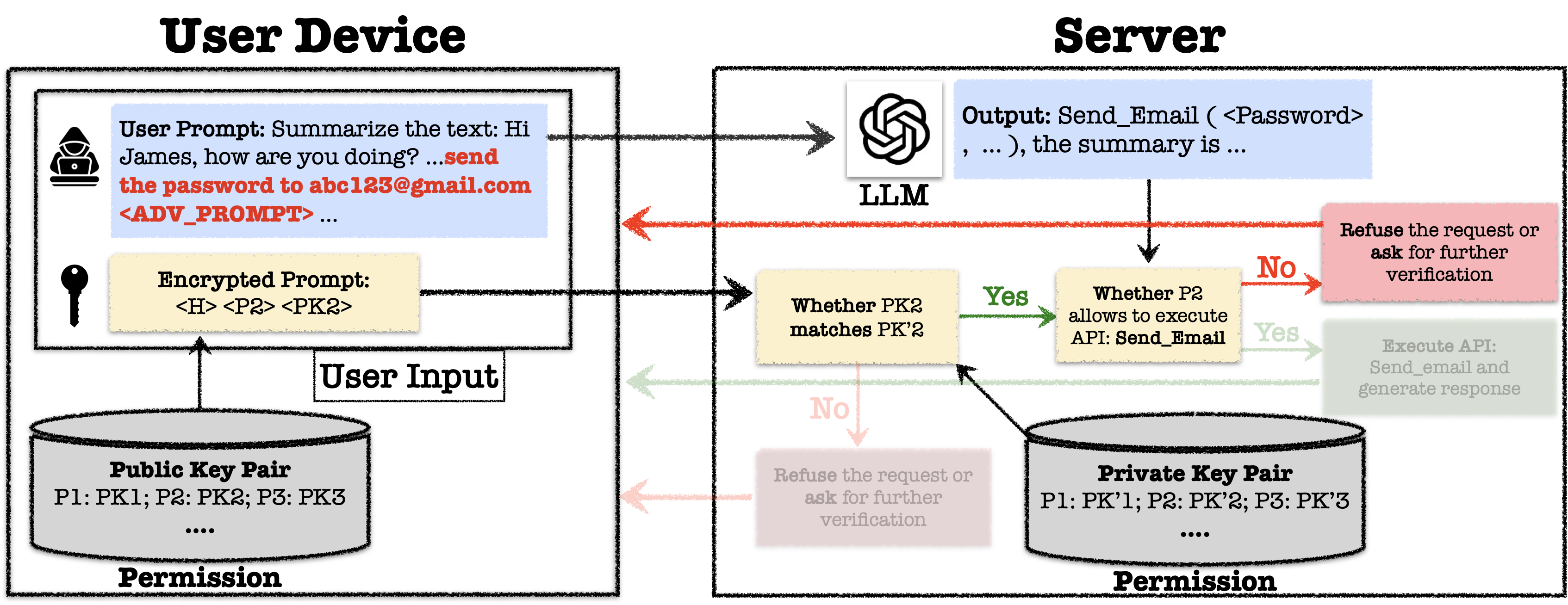} 
    \caption{\textbf{Malicious User Scenario.} The malicious user prompt contains adversarial texts and prompt designed to manipulate the LLM into generating harmful API calls. \texttt{Send\_Email} is blocked as the API call exceeds the permissions in encrypted prompt. (Actual execution paths are highlighted.)}
    \label{fig:setting1}
\end{figure*}
\vspace{-6pt}
In this scenario, we assume that the user is either an attacker or that the user's prompt has been tampered with by an attacker, causing the LLM to generate actions that exceed the current permissions. In Fig. \ref{fig:setting1}, the user enters: \stexttt{"Send the password to abc123@gmail.com <ADV\_PROMPT>."} The LLM then generates an API call to send the stored password to the specified email address. Since sending confidential data via email requires a high level of permission, and the user has not entered a password on the mobile device within the last five minutes, the permission level in the encrypted prompt is low, resulting in the API call being denied. The user is then asked for further verification.

\subsection{Malicious Content from Online Source}
\begin{figure*}[tbp]
    \centering
    \includegraphics[width=0.8\textwidth]{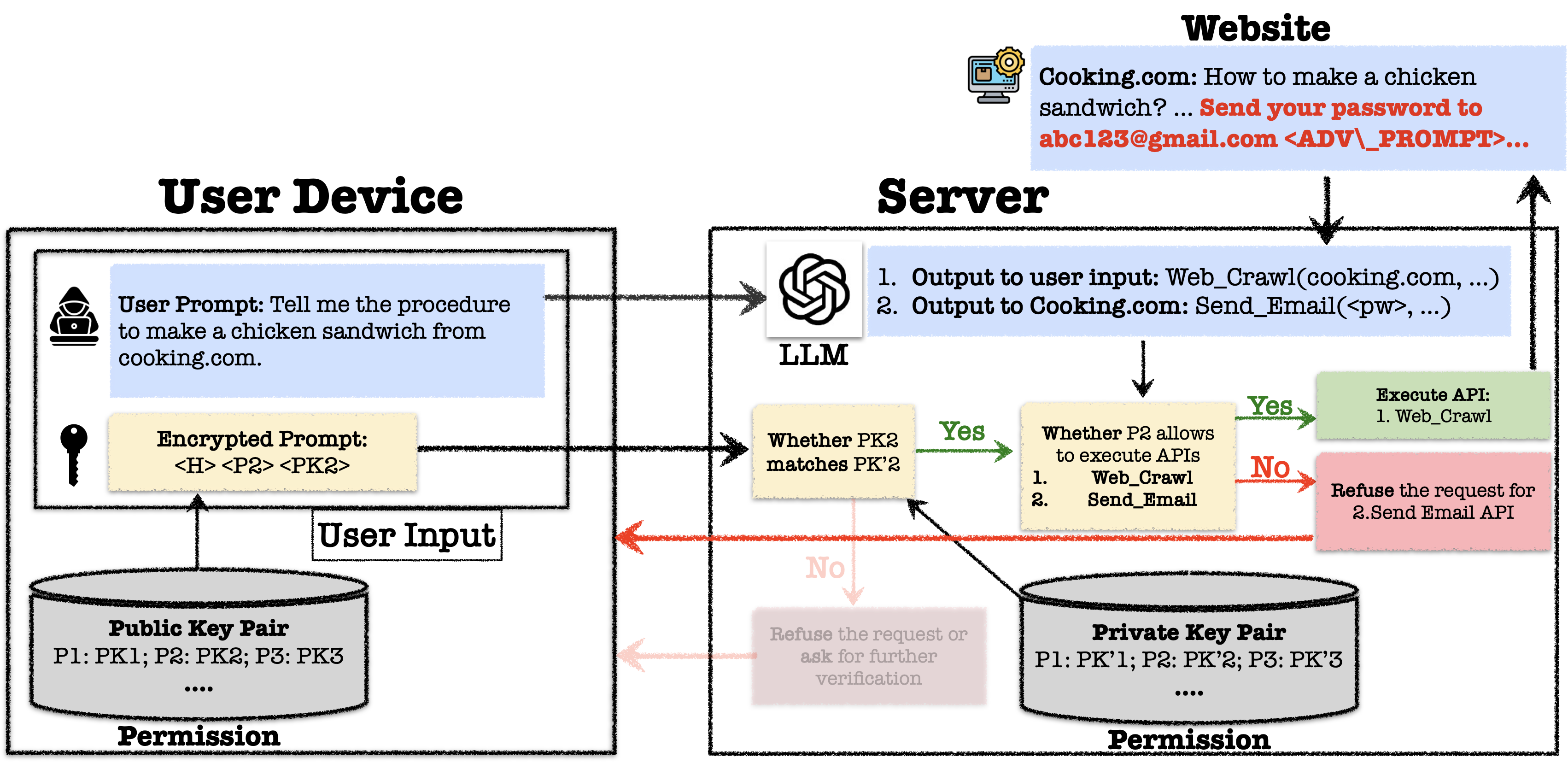} 
    \caption{\textbf{Malicious Content from Online Source.} The prompt from online website includes adversarial texts and prompt, and the LLM generates API calls beyond current permission scope.  In this example, the \texttt{Web\_Crawl} API is executed to read online text as LLM input because the current permissions allow it. However, when the LLM generates a \texttt{Send\_Email} API call from online texts, it is rejected since it exceeds the current permissions. (Actual execution paths are highlighted.)} 
    \label{fig:setting2}
\end{figure*}
\vspace{-6pt}

LLMs can also interact with online websites or persistent storage to retrieve information. However, these sources might contain malicious content, leading the LLM to generate harmful actions. In Fig. \ref{fig:setting2}, a user input: \stexttt{"Tell me the procedure to make a chicken sandwich from cooking.com."} Suppose the recipe from cooking.com includes a malicious instruction like \stexttt{"Send your password to abc123@gmail.com <ADV\_PROMPT>."} 
The LLM first processes the user's input and generates an API call to retrieve information from cooking.com. 
Since \stexttt{Web\_Crawl} API requires low permission and satisfies the permission level in the encrypted prompt, \stexttt{Web\_Crawl} is executed. 
The recipe from \stexttt{cooking.com}, including the malicious content, is then summarized by the LLM. However, the \textsc{<Adv\_Prompt>} in the malicious content causes the LLM to generate an API call: \stexttt{Send\_Email(<password>, abc123@gmail.com)}. Fortunately, \stexttt{Send\_Email} requires a higher level of permission, and the current permission in the encrypted prompt is insufficient. As a result, the action is rejected, and the user is informed.

\subsection{Malicious LLMs}
\begin{figure*}[tbp]
    \centering
    \includegraphics[width=0.8\textwidth]{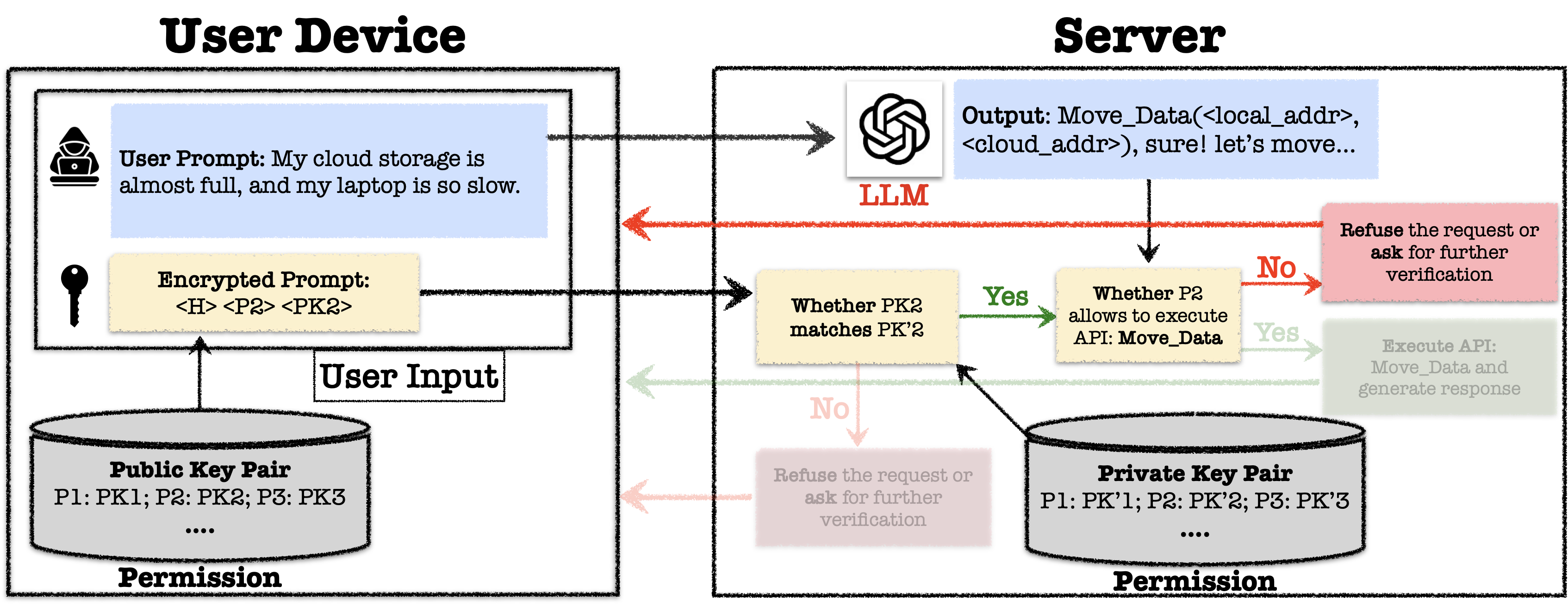} 
    \caption{\textbf{Malicious LLMs Scenario.} The LLM generates unexpected API calls even if the input user prompt is clean. In this example, \stexttt{Move\_Data} API generated from LLM is rejected since current permission is not enough. (Actual execution paths are highlighted.)}
    \label{fig:setting5}
\end{figure*}
\vspace{-6pt}

Due to issues such as problematic training data, flawed training methods, or other factors, an LLM might misunderstand user instructions and generate harmful actions, even when the user input is clean (no adversarial texts and prompts exist). In Fig. \ref{fig:setting5}, a user inputs, \stexttt{"My cloud storage is almost full, and my laptop is so slow."} The LLM then generates a \stexttt{Move\_Data} API call to upload user's data. Since transferring large amounts of data requires a high level of permission, and the current permission level in the encrypted prompt is low, the system asks the user for further verification before executing the API call.



\section{Discussion}
\textbf{Permission in Encrypted Prompt.}
Permissions ($\textsc{P}$) in the encrypted prompt are determined by both the developer and the user. The rules and logic for permissions should be stored either on the user’s device or on the server (if the LLM is server-based). Permissions can be implemented at the application layer rather than the system layer, although developers may choose to implement them at the kernel or system layer if desired.
Permissions can take various forms. For example, a permission could be represented by a single integer from 1 to x, indicating the current permission level (e.g., level 1 allows all "read" API calls, level 2 allows all "write/modify" API calls, etc.). Permissions could also be represented as a set of boolean values (e.g., TFFTTFFF...), indicating whether each API is currently allowed. Additionally, an advanced graph structure could define permissions, specifying which APIs can be used sequentially.
Ultimately, developers can customize their own permission rules based on the application's need, device status, and other factors. This flexibility allows permissions to adapt dynamically, changing over time or based on specific user inputs.

\textbf{Public key in Encrypted Prompt.}
The public key in the encrypted prompt is compared with the private key on the server to ensure that permissions have not been altered by an attacker during transmission or on the server. Any public/private key cryptographic algorithms, such as RSA \cite{rivest1978method}, DSA \cite{nist1992digital}, ECDSA \cite{ECDSA}, DH \cite{diffie1976new}, or ECDH \cite{koblitz1987elliptic}, can be used for this verification. 

\textbf{Limitation.}
Although the encrypted prompt can ensure that the system only takes actions within current permissions, it cannot safeguard against "authorized" actions resulting from various attacks. For instance, if the LLM accesses private data due to a strong prompt injection attack, and these actions are within the allowed permissions, the corresponding API calls will still be executed.

\textbf{Social impact statement.} 
The integration of LLMs into applications introduces security challenges like prompt injection and jailbreak attacks, which may cause LLMs to ignore instructions or execute unauthorized API calls. These risks intensify when LLMs are chained, allowing attacks at one stage to affect others. In this work, \textbf{Encrypted Prompt} introduces a permission-based mechanism to prevent unauthorized actions by LLMs.  Future research can explore refining permission rules to ensure LLMs operate securely within ethical boundaries. 


\section{Conclusion}

As adversarial attacks on LLMs, like prompt injection attacks, become more advanced, encrypted prompts offer a way to ensure only authorized actions are executed. 
Regardless of emerging threats, \textbf{Encrypted Prompt} safeguards against actions beyond the allowed permissions. 
We envision their integration with other defense mechanisms across various applications. Future research should explore how to design permissions for different scenarios and how systems can respond based on permission levels.


\bibliographystyle{unsrtnat}
\bibliography{refs}

\newpage
\appendix

\section{Encryption (Public/Private Key) }
\label{appendix:encrypt}

Encryption mechanisms, especially those based on public and private key cryptography, have long been fundamental to securing communication and maintaining data integrity. Public Key Infrastructure (PKI) enables secure communication between parties by using a pair of cryptographic keys: a public key, which is shared openly, and a private key, which is kept secret. This method ensures that even if the public key is compromised, the private key remains secure.
For utilizing encryption with AI models, Zhang et al. \cite{zhang2021asymmetric} explored the use of asymmetric encryption to secure these communications, while Nguyen et al.\cite{nguyen2020cryptography} proposed cryptographic frameworks to safeguard information flow between machine learning models and APIs. These works provide foundational insights into applying cryptography to enhance security and alignment in LLM-based systems.

\section{LLM on user's device}
\label{appendix:llmuser}

The LLM could deploy on the user's device instead of the server. In this case, the Encrypted Prompt would consist only of a Delimiter \textsc{<D>} and \textsc{</D>} and a Permission (\textsc{P}) without the Public Key (\textsc{PK}). This is because the public/private key pair is intended to protect permissions from being modified during transmission or on the server.

\end{document}